# High-Pressure Synthesis of a Pentazolate Salt

Brad A. Steele,[†] Elissaios Stavrou,[*,‡] Jonathan C. Crowhurst,[‡] Joseph M. Zaug,[‡] Vitali B. Prakapenka,[¶] and Ivan I. Oleynik[*,†]

*Department of Physics, University of South Florida, 4202 East Fowler Ave., Tampa, FL 33620, Lawrence Livermore National Laboratory Livermore California, Physical and Life Sciences Directorate, P.O. Box 808, Livermore, California 94550, USA, and Center for Advanced Radiation Sources, University of Chicago, IL 60637, USA*

E-mail: stavrou1@llnl.gov; oleynik@usf.edu

## Abstract

The pentazolates, the last all-nitrogen members of the azole series, have been notoriously elusive for the last hundred years despite enormous efforts to make these compounds in either gas or condensed phases. Here we report a successful synthesis of a solid state compound consisting of isolated pentazolate anions $N_5^-$, which is achieved by compressing and laser heating cesium azide ($CsN_3$) mixed with $N_2$ cryogenic liquid in a diamond anvil cell. The experiment was guided by theory, which predicted the transformation of the mixture at high pressures to a new compound, cesium pentazolate salt ($CsN_5$). Electron transfer from Cs atoms to $N_5$ rings enables both aromaticity in the pentazolates as well as ionic bonding in the $CsN_5$ crystal. This work provides a critical insight into the role of extreme conditions in exploring unusual bonding routes that ultimately lead to the formation of novel high nitrogen content species.

---

[*]To whom correspondence should be addressed
[†]Department of Physics, University of South Florida, 4202 East Fowler Ave., Tampa, FL 33620
[‡]Lawrence Livermore National Laboratory Livermore California, Physical and Life Sciences Directorate, P.O. Box 808, Livermore, California 94550, USA
[¶]Center for Advanced Radiation Sources, University of Chicago, IL 60637, USA

# Introduction

Although most of heterocyclic compounds have been synthesized in the second half of the 19th century, cyclo-$N_5$, the last elusive member in the azole series, was discovered only in the mid-1950s as a part of an aryl pentazole molecule[1,2]. This all-nitrogen analog of cyclopentadienyl anion was proven to be aromatic with N-N bond lengths $1.3 - 1.35$ Å, intermediate between single (hydrazine, 1.45 Å) and double (*trans*-diimine, 1.25 Å) bonds[3]. Numerous attempts to isolate the pentazole ($HN_5$) or the $N_5^-$ anion[4–9] were unsuccessful until Vij *et al* produced $N_5^-$ in the gas phase in 2002 by cleaving the C-N bond in substituted phenylpentazoles[10]. This was followed by an independent confirmation by Ostmark *et al* in 2003[11]. Most recently, $N_5^-$ was prepared in a THF solution and was reported to be stable below $-40°C$[12]. The enormous interest in these pentazolates has been driven by the possibility of releasing a large amount of energy upon conversion of the single-double N-N bonds in the $N_5^-$ aromatic ring to triple N-N bonds in the $N_2$ molecule. This makes pentazolates potentially important components for the development of new, green, high-energy-density materials.

Being one of several metastable nitrogen species, pentazolates provide an excellent opportunity to bridge the gap between a wealth of theoretically predicted exotic molecules containing only N atoms and the experimental difficulties for their synthesis[10,13–15]. Besides well-studied azides containing double-bonded linear $N_3^-$ anions[16–18], gas phase pentazolate $N_5^-$ anions[10], $N_5^+$ chain cations[13], and gas phase tetranitrogen $N_4$[19], the only other experimentally observed all-nitrogen compounds are non-molecular crystalline phases of nitrogen, which were synthesized at very high pressures ($> 120\,\text{GPa}$) and temperatures ($> 2000\,\text{K}$)[15,20,21]. These polymeric cubic-gauche (cg)[15,20] and layered[21] phases of nitrogen are metastable at pressures down to $40\,\text{GPa}$, however, their quenching to ambient pressure and temperature is problematic. Our recent calculations[22] predicted the thermodynamic stability of sodium pentazolate $NaN_5$ at high pressures, as well as indicated a possibility that such a compound might have been synthesized by Eremets *et al* in 2004[18]. Therefore, the exploration of solid-phase pentazolates[22–26] can provide a viable route towards extending the range of metastability of high nitrogen compounds with the ultimate goal of their recovery at ambient conditions.

Here we report the first successful synthesis of pentazolates in the solid phase, which is achieved by compressing and laser heating cesium azide ($CsN_3$) mixed with $N_2$ cryogenic liquid in a diamond anvil cell (DAC). The experiment was guided by theory, which predicted the transformation of the mixture at high pressures to a new compound, cesium pentazolate salt ($CsN_5$). In addition, the calculations explain why an alkali metal is the key to achieving enhanced stability of pentazolates in the solid phase: highly electron-donating cesium transfers maximum negative charge to the $N_5$ rings , thus enabling both aromaticity in the isolated $N_5^-$ and ionic bonding between cesium cations and pentazolate anions within the crystalline environment. The synthesis of $CsN_5$ is monitored by visual observations, synchrotron X-ray diffraction (XRD), and confocal Raman spectroscopy. The choice of cesium over lighter alkali cations is dictated by the large scattering factor of Cs, which is proportional to the number of the electrons per cation. The complimentary Raman spectroscopy is used to uncover features associated with the covalent bonding configuration within the pentazolates $N_5^-$, which contribute insignificantly to the observed XRD patterns.

## Methods

**Experimental:** Commercially available (Sigma-Aldrich) 99.99% pure $CsN_3$ is placed inside an argon gas purged glovebox and ground to fine powder. The fine powder sample with incorporated pressure sensors is loaded into a diamond-anvil cell (DAC). The DAC sample cavity is filled cryogenically with nitrogen which serves both as a pressure transmitting medium and reagent. The sample chamber is sealed and the pressure is increased to the target value. Special care is taken to achieve as uniform mixture of $CsN_3$ and $N_2$ as possible. Pressure is determined using known ambient temperature EOS of gold[27] and calibrated ruby luminescence[28]. A CCD detector is used to collect pressure dependent X-ray diffraction (XRD) data at the Advanced Photon Source GSE-CARS (sector 13, $\lambda = 0.3344$ Å or $\lambda = 0.310$ Å)[29] and at the Advanced Light Source Beamline 12.2.2 ($\lambda = 0.4959$ Å)[30]. The monochromatic x-ray beams were focused to a nominal 4-10 $\mu$m diameter spot. Raman studies are performed using the 514 nm line of an argon-ion laser in the backscattering geometry, which is focused to a 2 $\mu$m probe diameter. Raman spectra are collected with a spectral resolution of 4 cm$^{-1}$ using a single-stage dispersion grating spectrograph equipped with a CCD array detector. Laser heating is performed in a double-sided laser heating geometry. Temperature is determined spectroradiometrically. Although CsN$_3$ is optically transparent at ambient pressure, band closure with increasing pressure results in an opaque sample above 30 GPa, which allows for extremely efficient optical coupling with near-infrared laser irradiation.

**Theoretical:** The crystal structure search for new Cs$_x$N$_y$ compounds is performed using the first principles evolutionary structure prediction method USPEX[31–33]. The convex hull is first constructed during variable composition search at several pressures, 0.5 GPa, 30 GPa, and 60 GPa, followed by fixed stoichiometry crystal structure search using up to 8 formula units per unit cell. First-principles calculations are performed by employing the Perdew-Burke-Ernzerhof (PBE) generalized gradient approximation (GGA)[34] to density functional theory (DFT) as implemented in VASP DFT package[35]. Projector augmented wave (PAW) pseudopotentials[36] and plane wave basis set with an energy cutoff of 600 eV and a 0.05 Å$^{-1}$ k-point sampling of the Brillouin zone are used for structure search. To describe a substantial overlap of electron density between nitrogen atoms in triply-bonded N$_2$, a hard pseudopotential for N and correspondingly higher cutoff of 1000 eV is used to produce the accurate convex hulls, shown in Figure 1 (a). The dispersive correction due to Grimme[37] is added to DFT energy and forces to take into account the long-range van der Waals forces, which are significant in the system under investigation due to the large polarizability of Cs atoms. An adequate accuracy of these calculations is established by comparing the formation enthalpies calculated by the PBE GGA functional with vdW with the standard PAW N pseudopotential and the HSE06 hybrid functional with the hard N pseudopotential. The vibrational properties are calculated within the frozen phonon approximation. The off-resonant Raman frequencies and corresponding activities (intensities) are obtained by computing phonon modes at the Γ-point together with the corresponding derivatives of the macroscopic dielectric tensor with respect to the normal mode coordinates[38,39].

# Results and discussion

To guide experimental synthesis, we performed an extensive first-principles structural search for Cs$_x$N$_y$ compounds of varying stoichiometry at several pressures using the evolutionary structure prediction method USPEX[31–33]. The multiple crystal structures discovered during the search are classified by constructing the convex hull – the formation enthalpy-composition curves, each corresponding to a given pressure (Figure 1(a)). The points connected by segments represent the lowest enthalpy structures at a specific composition and pressure. The formation enthalpy is the difference between the enthalpy of the particular crystal and the enthalpy of the elemental compounds. The reference structures are the $\alpha$-$N_2$, $\epsilon$-$N_2$, and cg-N[20,40] crystal phases for nitrogen, and the bcc-Cs and *Cmca*-Cs (Phase V) phases for Cs[41], each taken at its corresponding pressure of stability. The structures and stoichiometries on the convex hull are thermodynamically stable, while those above the convex hull are metastable.

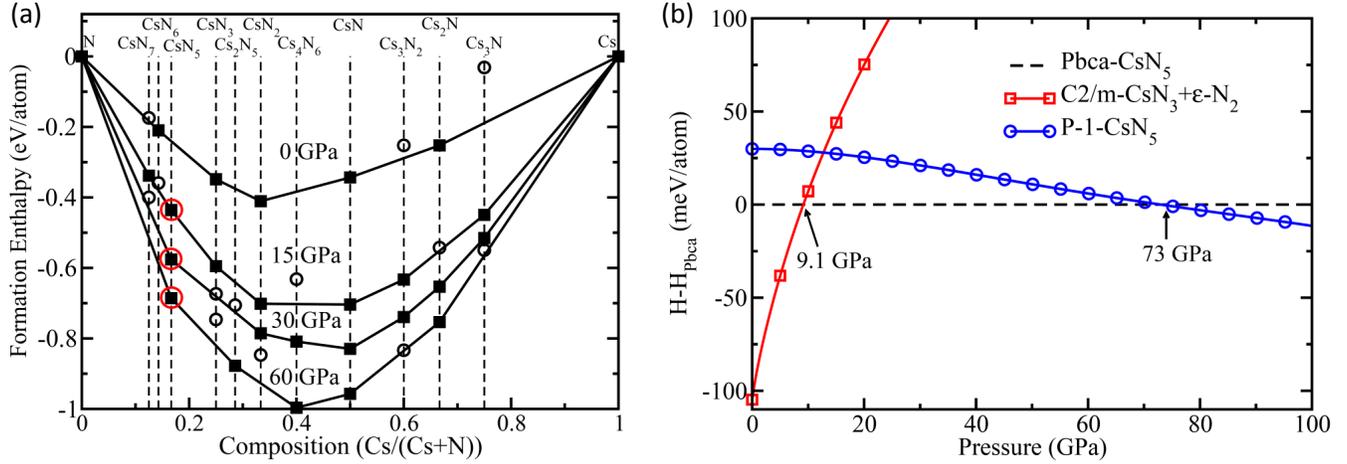

Figure 1: (a) Convex hull diagram at pressures up to 60 GPa. Solid squares represent stable phases, open circles – metastable phases, large red circles – stable $CsN_5$ phase on the convex hulls. (b) Relative enthalpy difference between two cesium pentazolate ($CsN_5$) polymorphs and the $\varepsilon$-$N_2$ phase of solid nitrogen plus the $C2/m$-$CsN_3$ phase of cesium azide as a function of pressure.

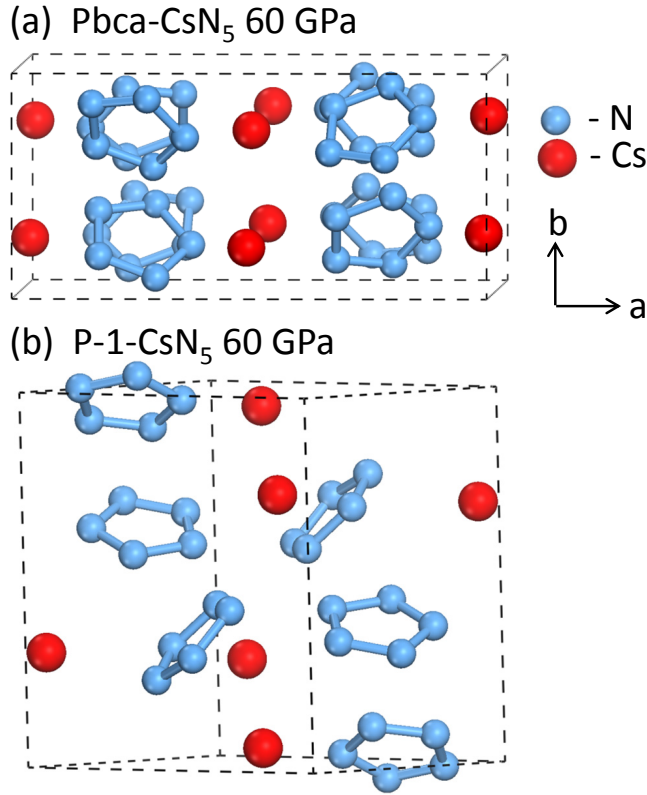

Figure 2: Crystal structures of two energetically competitive cesium pentazolate ($CsN_5$) polymorphs at 60 GPa discovered during the structure search: (a) $Pbca$-$CsN_5$; (b) $P$-$1$-$CsN_5$.

The calculations demonstrate that cesium pentazolate salt $CsN_5$, consisting of pentazolate anion rings ($N_5^-$) and cesium cations, is on the convex hull at a surprisingly low pressure of just 15 GPa (Figure 1(a)). The relative enthalpy difference between the mixture of cesium azide plus di-nitrogen ($CsN_3+N_2$) and reference pentazolate phase *Pbca*-$CsN_5$, plotted in Figure 1(b), indicates that $CsN_5$ is energetically preferred above a relatively low pressure of 9.1 GPa. The latter pressure is about 41 GPa lower than the predicted pure nitrogen transition pressure to the single-bonded 3-fold coordinated cubic gauche phase of nitrogen (cg-N)[20]. Therefore, the synthesis of $CsN_5$ requires much lower pressure and temperature stimuli than is needed to synthesize cg-N (over 100 GPa and 2,000 K)[15].

The structure search found two energetically competitive polymorphs of cesium pentazolate salt: one with space group *Pbca* (*Pbca*-$CsN_5$) and the second – with space group *P-1* (*P-1*-$CsN_5$) (Figure 2(a) and (b) respectively). The lowest enthalpy polymorph at 60 GPa is Pbca-$CsN_5$, which contains eight formula units in the unit cell, while the energetically competitive *P-1*-$CsN_5$ polymorph – six. *P-1*-$CsN_5$ is only 6 meV/atom higher in enthalpy than *Pbca*-$CsN_5$ at 60 GPa, and is the lowest enthalpy polymorph above 73 GPa (Figure 1(b)). Both structures consist of layers of Cs atoms sandwiched by layers of pentazolate $N_5^-$ anions oriented differently in the unit cell with respect to one another (Figure 2(a) and (b)). In both crystals there are two layers of cyclo-$N_5^-$ in the unit cell: the *P-1*-$CsN_5$ layer contains three cyclo-$N_5^-$, while the *Pbca*-$CsN_5$ layer – four pentazolates. The somewhat complex arrangement of pentazolates in the unit cell for *P-1*-$CsN_5$ reduces the symmetry to P-1.

Due to the aromatic nature of the bond in the pentazolate $N_5^-$ ring, the bond lengths for both phases are between those of the double-bond (1.25 Å as in *trans*-diimine ), and the single bond (1.45 Å as in hydrazine). All eight cyclo-$N_5^-$ in the unit cell of the Pbca-CsN$_5$ crystal have approximately the same bond lengths, e.g. 1.328 Å at 0 GPa, see Table 1. The average bond length in the *P-1*-CsN$_5$ crystal is also 1.328 Å, see Table 1, however, the 0.01 Å spread of the five bond lengths around their average is slightly larger than the 0.005 Å spread in the *Pbca*-CsN$_5$ structure, which correlates with the reduced symmetry of the *P-1*-CsN$_5$ crystal. The aromaticity of $N_5^-$ is further quantified by calculating the Mayer bond orders[42,43], which are found to be ~1.4 for both phases, i.e. between single and double bonds, see Table 1. Both phases exhibit a significant amount of electron transfer from Cs atoms to N$_5$ rings giving each pentazolate a total charge of approximately -0.85, the average charge per atom being -0.167, see Table 1. The two CsN$_5$ polymorphs exhibit slightly different charge distributions and bond orders due to the different crystal structures and slight distortions of the $N_5^-$ rings.

Table 1: Calculated average Mulliken charges, Mayer bond orders, and bond lengths in $N_5^-$ rings of CsN$_5$ compounds, which are compared to those in N$_3$ chains of cesium azide at 0 GPa.

| Crystal | N-cluster | Charges (e/atom) | Bond Order | Bond Length (Å) |
|---|---|---|---|---|
| *P-1*-CsN$_5$ | N$_5$ ring | -0.169 | 1.407 | 1.328 |
| *Pbca*-CsN$_5$ | N$_5$ ring | -0.167 | 1.420 | 1.328 |
| *I4/mcm*-CsN$_3$ | N$_3$ chain | -0.230 | 1.96 | 1.188 |

To synthesize the newly predicted cesium pentazolate salt, cesium azide is loaded into a DAC together with N$_2$ cryogenic liquid, compressed to high pressures, and then laser heated. The XRD patterns and pressure-dependent Raman spectra before laser heating are given in Supporting Information, see Figures S6 and S7 respectively. Laser-heating the CsN$_3$+N$_2$ mixture at ca 60 GPa results in a remarkable change in both the XRD pattern and the Raman spectrum of the quenched compound. Visual observation of the DAC encapsulated material indicates the presence of a transparent material exactly at the laser heated spot as shown in the inset to Fig. 3(a). Examination of XRD patterns given in the Supporting Information, Figure S8, reveal the appearance of new intense and narrow Bragg peaks with a simultaneous disappearance of the broad $CsN_3$ peaks. The Bragg peaks of the new phase are well indexed using the theoretically predicted P-1-$CsN_5$ structure. Preferred orientation (spotty Debye rings) and strongly anisotropic peak broadening effects (Figure S7), which are normal in HP-HT synthesis, prevent us from a full structural refinement (Rietveld) of the positional parameters. The resultant Le Bail-fitted and experimental XRD patterns at 60 GPa, shown in Fig. 3(a), are well-matched. The lattice parameters obtained from the Le Bail refinement are close to those predicted by theory, see Table 2. In particular, the experimental and theoretical unit cell volumes differ by less than 5 %, which is within the expected accuracy of DFT calculations[44].

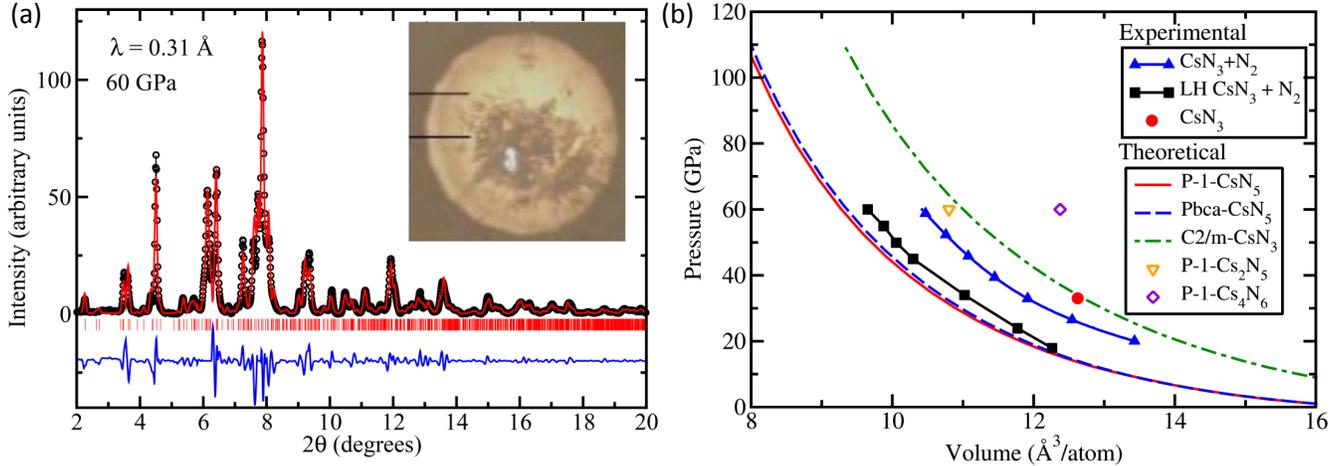

Figure 3: (a) Measured X-Ray diffraction patterns (black circles) after laser heating against the Le Bail fit (red solid line) using the predicted *P-1*-$CsN_5$. Vertical red ticks mark positions of Bragg peaks. The blue line is the difference between the measured and fitted intensities. The inset shows a microphotograph of the sample after heating in transmitted light, indicating the transparency of the synthesized phase. (b) Comparison of the measured and calculated pressure versus volume equation of state (EOS) for the synthesized compound: experimental data are plotted with solid symbols, whereas theoretical predictions – by open symbols and lines. Also shown are the experimental (P,V) points corresponding to $CsN_3$ at 33 GPa and the EOS of $CsN_3+N_2$ mixture before laser heating (LH).

The comparison between the experimental pressure versus volume equation of state (EOS) with the theoretical EOS of $CsN_5$ and various candidate structures obtained during the evolutionary structure search is shown in Figure 3(b). The experimental EOS matches the theoretical EOSs well for both P-1-$CsN_5$ and Pbca-$CsN_5$ phases in the full range of applied pressures. In addition, the measured EOS is compared to the theoretical EOS of other crystals also predicted to be stable at

Table 2: Comparison between the theoretical lattice parameters of P-1-CsN$_5$ and those obtained from the experimental refinement at 60 GPa.

| Quantity | Exp. Refinement | Theory | Difference |
|---|---|---|---|
| a(Å) | 6.784 | 6.703 | -1.19 % |
| b(Å) | 7.884 | 7.767 | -1.48 % |
| c(Å) | 6.581 | 6.552 | -0.44 % |
| $\alpha$ | 85.136° | 80.689° | -5.22 % |
| $\beta$ | 94.344° | 95.115° | 0.81 % |
| $\gamma$ | 89.934° | 85.421° | -5.01 % |
| V(Å$^3$) | 349.75 | 333.74 | -4.58 % |

60 GPa, Cs$_4$N$_6$ and Cs$_2$N$_5$. However, as can be seen in Figure 3(b), the calculated volumes of these structures are significantly higher than those of CsN$_5$ in the same pressure range. In addition, the experimental volume of CsN$_3$+N$_2$ after laser heating is 10% less than the sum of volumes of CsN$_3$ and N$_2$ before laser heating at a given pressure. This provides a strong indication of major atomic rearrangements to form N$_5$ rings during the conversion of CsN$_3$+N$_2$ to CsN$_5$.

As evidenced from XRD measurements (Supporting Information, Figure S10) and optical observations, the newly synthesized and optically transparent phase of CsN$_5$ remains stable down to 18 GPa. Below this pressure the disappearance of the crystalline pattern from XRD measurements indicates the formation of an amorphous substance. The optical transparency of the material in the laser-heated spot, indicating a wide band gap material, is another confirmation of the synthesis of a CsN$_5$ compound. All other predicted stable and metastable compounds are calculated to be either metallic or small-band-gap semiconductors ($E_g < 1.3$ eV) in the same pressure range. Hence, they would be optically opaque, if observed. The calculated band structure of the P-1 phase of CsN$_5$ (Supporting Information, Figure S3) confirms that cesium pentazolate is an optically transparent insulator with a band gap of 3.2 eV at 60 GPa.

The Raman spectra obtained after laser heating reveal the disappearance of the Raman modes of the azide anion N$_3^-$: sharp peaks $\nu_2$ at 656 cm$^{-1}$, $\nu_1$ at 1,490 cm$^{-1}$, and the broad peak in the 1,750-2,000 cm$^{-1}$ frequency range (Figure 4(a) and Supporting Information, Figure S11). The $\nu_1$ mode is a shoulder to the first-order diamond Raman peak (from the diamond culets), which is also seen in the Raman spectra of $CsN_3+N_2$ taken at increasing pressures upon compression before laser heating (Supporting Information, Figure S7). New Raman peaks appear between 700 and 850 cm$^{-1}$, and at 1,170 cm$^{-1}$, and 1,320 cm$^{-1}$, the latter being seen only below 60 GPa due to the screening by the diamond-anvil Raman peak at higher pressures (Supporting Information, Figure S11). Both the relative intensities and the frequencies of the experimentally observed Raman peaks are in agreement with those of the theoretically predicted Raman modes of P-1-$CsN_5$ (Figure 4(a)). The observed Raman modes at 55 GPa are assigned as follows: 1,320 cm$^{-1}$ – $N_5^-$ ring breathing; 1,170 cm$^{-1}$ – $N_5^-$ antisymmetric N-N breathing and angle deformations, and 700-850 cm$^{-1}$ – $N_5^-$ ring bending and librational modes.

Table 3: Raman mode assignments and corresponding experimental and theoretical Raman frequencies and pressure-dependent slopes $(\partial\omega/\partial P)_T$. The frequencies and slopes are reported at two different pressures: the high frequency modes – at 35 GPa (the diamond-anvil Raman peak shields sample peaks at higher pressures); the low frequency $N_5^-$ bending doublet – at 40 GPa (these modes are not clearly resolved at lower pressures).

| Raman mode assignment | Frequency (Exp.) cm$^{-1}$ | Frequency (Calc.) cm$^{-1}$ | Pressure slope $(\partial\omega/\partial P)_T$ (Exp.) cm$^{-1}$GPa$^{-1}$ | Pressure slope $(\partial\omega/\partial P)_T$ (Calc.) cm$^{-1}$GPa$^{-1}$ |
|---|---|---|---|---|
| Anti-symmetric $N_5^-$ breathing | 1270 (35 GPa) | 1338.5 (35 GPa) | 2.88 | 2.13 |
| Symmetric $N_5^-$ breathing | 1246 (35 GPa) | 1287.2 (35 GPa) | 3.19 | 1.80 |
| $N_5^-$ anti-symmetric breathing and deformation | 1158 (35 GPa) | 1121.4 (35 GPa) | 0.43 | 0.75 |
| $N_5^-$ bend | 785 (40 GPa) | 786.2 (40 GPa) | 0.62 | 0.55 |
| $N_5^-$ bend and libration | 716 (40 GPa) | 762.5 (40 GPa) | 0.76 | 0.07 |
| $N_5^-$ libration and lattice modes | 310 (35 GPa) | 308.2 (35 GPa) | 2.71 | 2.92 |

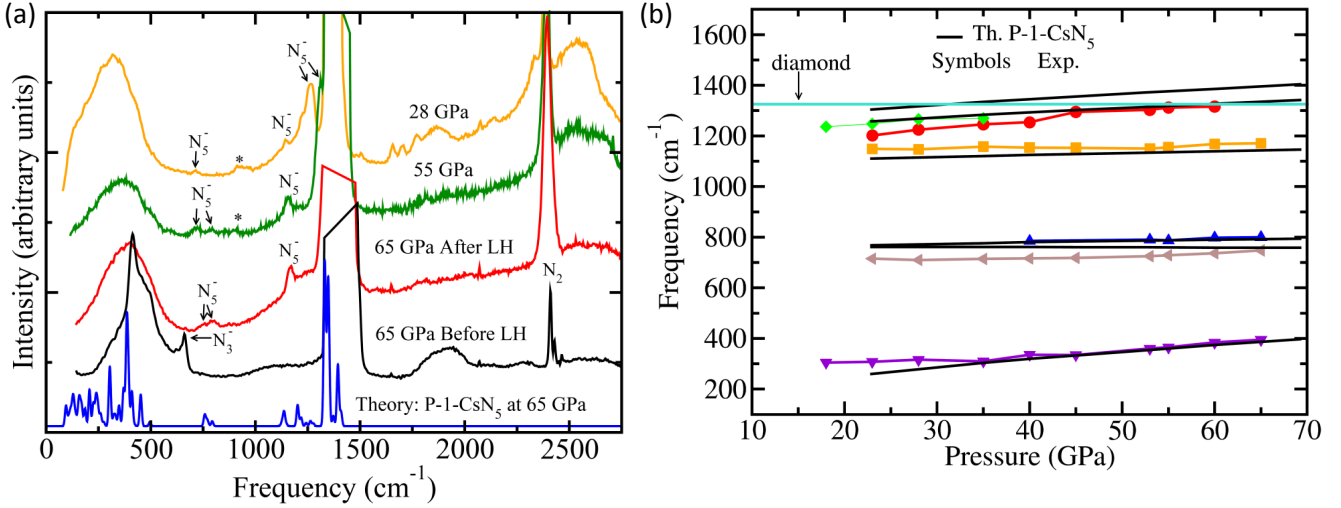

Figure 4: (a) Experimental Raman spectra of the $CsN_3+N_2$ mixture at 65 GPa before laser heating (LH) (black curve), and of $CsN_5$ after LH (red curve) then upon decompression to 55 GPa (green curve) and 28 GPa (amber curve). Theoretical Raman spectrum of $P\text{-}1$-$CsN_5$ polymorph (blue curve) is also shown for comparison. Arrows indicate $N_5^-$ peaks while the asterisk identifies an artifact of the measurements, see discussion in the text. (b) Comparison of pressure dependence of experimental and theoretical frequencies of Raman-active modes of $CsN_5$: theory - black lines, experiment - colored symbols connected by different colored lines for different modes. The light blue line represents the lower frequency of the first order Raman scattering of diamond.

Mode assignments for each experimentally measured mode as well as corresponding frequencies and pressure-dependent slopes $(\partial\omega/\partial P)_T$ are compared with those calculated by theory and given in Table 3. The comparison of the pressure dependencies of measured and predicted Raman frequencies (Figure 4(b)) shows very good overall agreement, both decreasing smoothly upon pressure reduction.

A broad Raman feature appearing in the low frequency region around 300 cm$^{-1}$ is probably the convolution of the multiple closely-spaced lattice modes that are broadened due to local non-hydrostatic conditions in the sample's chamber. Its experimental frequency is determined as the maximum of this bump and is plotted as a function of pressure in Figure 4(b) and reported in Table 3. The calculated frequency of this "average" lattice mode is determined by selecting the most intense Raman peak in this frequency interval. Additional closely spaced Raman bands that appear in the calculated Raman spectrum are not observed in the experimental data due to their weak intensity and close proximity to stronger intensity bands. Therefore, these additional modes are not plotted in Figure 4(b), but can be seen in Figure 4(a) as well as Figure S5.

The $N_5^-$ bending doublet modes (Table 3) that appear in the frequency interval 700–850 cm$^{-1}$ have a very weak intensity. The reduced signal-to-noise ratio in this spectral region overlaps with the high fluorescence background present in the samples synthesized at HP-HT conditions, which is specifically seen at 55 GPa. To address this challenge, the $N_5^-$ bending doublet modes are identified as those that consistently appear at various pressures, and at a given pressure have the same frequencies regardless of various probing points of the sample. This procedure gives us two clear modes in this frequency interval 700–850 cm$^{-1}$, which can be traced down to 40 GPa. At lower pressures one mode disappears, and the other is traced down to 20 GPa, see Figures 4(a) and 4(b). This observation matches well with the trends predicted by theory: as seen from Figure S5, the intensity of these modes substantially decreases with pressure. These modes are marked by arrows in Figure 4(a). There is an additional feature in the Raman spectra in this frequency interval measured upon decompression, which does not consistently appear at each pressure and does not display any pressure dependence. We conclude that this random signal arises from the broad high intensity fluorescence background and thus it is marked with an asterisk in Figure 4(a) to identify this artifact.

Upon pressure release, the intensity of the Raman modes of $CsN_5$ drops substantially, and only the most intense internal $N_5$ mode (1,236 cm$^{-1}$) and the intense broad feature at the low frequency interval (0-550 cm$^{-1}$) are observed below 20 GPa (4(b)). Upon further release of pressure below 18 GPa, all Raman modes other than the one corresponding to high frequency nitrogen vibrons disappear. The N$_2$ vibrons are always present because nitrogen is the pressure transmitting medium. Thus, XRD, Raman measurements and optical observations indicate the disappearance of crystalline CsN$_5$ below 18 GPa. As no signal from any crystalline substance is observed, we conclude that the CsN$_5$ compound amorphizes below this pressure, rather than decomposes to elemental compounds. This trend has also been observed in other materials synthesized at high pressures[45].

It is worth mentioning that a similar Raman spectrum was observed upon compression and laser heating of sodium azide NaN$_3$ by Eremets *et al*[18], which was later interpreted as an appearance of the new compound, sodium pentazolate NaN$_5$, containing N$_5^-$ anions[22]. However, due to the limited amount of experimental data, the comparison made between theory and experiment was less than satisfactory, which did not allow for a definite conclusion about the synthesis of the new compound in the experiments of Eremets *et al*[18]. Additional details on this comparison as well as discussion of other related work are provided in the Supporting Information.

# Conclusion

In summary, we have successfully synthesized cesium pentazolate salt, the first solid state compound consisting of pentazolate anions N$_5^-$. Guided by theory, CsN$_5$ was successfully synthesized in a DAC by compressing and laser heating a CsN$_3$ and N$_2$ precursors at pressures near 60 GPa. The chemical transformation to CsN$_5$ is visually observed by the appearance of an optically transparent wide band gap area at the laser-irradiated spot. The refinement of the XRD pattern using the space group information provided by theory unambiguously proves that the crystal structure of the new compound contains 6 formula units per unit cell and has a space group symmetry of P-1. Raman-active mode assignments of experimental Raman spectra assisted by the theory provide convincing fingerprints of the N$_5^-$ anion. This work demonstrates that high-pressure/high-temperature experiments guided by first-principles crystal structure prediction is a powerful approach for discovering condensed phase compounds not easily achievable by using standard tools of synthetic chemistry. Moreover, this work paves the way for the synthesis of novel, high energy density materials with the promise of quenching metastable phases at ambient conditions.

## Supporting Information Available

Snapshots of additional predicted $Cs_xN_y$ structures, electronic band structure, convex hull calculated with HSE, additional XRD and Raman spectra, experiments with Argon  This material is available free of charge via the Internet at `http://pubs.acs.org/`.

## Acknowledgement

This research is supported by Defense Threat Reduction Agency, grant HDTRA1-12-1-0023. This work was also performed under the auspices of the U. S. Department of Energy by Lawrence Livermore National Security, LLC under Contract DE-AC52- 07NA27344. GSECARS is supported by the U.S. NSF (EAR-0622171, DMR-1231586) and DOE Geosciences (DE-FG02-94ER14466). Use of the APS was supported by the DOE-BES under Contract No. DE-AC02-06CH11357. The Advanced Light Source is supported by the Director, Office of Science, Office of Basic Energy Sciences, of the U.S. Department of Energy under Contract No. DE-AC02-05CH11231.

## Corresponding Authors

email: oleynik@usf.edu and stavrou1@llnl.gov.

## Notes

The authors declare no competing financial interests.